

 \magnification=1200
\nopagenumbers
\vfill
\line{\hfil UB-ECM-PF-94/12}
\line{\hfil June 1994}
\vskip 2truecm
\font\gran=cmbx12
{ \centerline{\gran Some Remarks on the Matching Conditions }}
\vskip 1truecm
\bigskip
\centerline{D. Espriu  ~and~  J. Matias}
\bigskip
\centerline{\it D.E.C.M., Facultat de F\'\i sica and I.F.A.E.}
\centerline{\it Universitat de Barcelona}
\centerline{\it Diagonal, 647}
\centerline{\it E-08028 Barcelona}
\vskip 1.5truecm
\centerline{ABSTRACT}
 \bigskip
We analyze the matching conditions
to determine the values that the ${\cal O}(p^4)$ coefficients of an
Effective Chiral Lagrangian take in the Standard Model in the
limit of a large Higgs mass, pointing out a number of subtleties
that appear to have gone unnoticed previously. We apply
the resulting Effective Chiral Lagrangian,
including the leading two loop effects, to analyze the
most recent electroweak data assuming $m_{top}=174\pm 10$ GeV.

\vfill
\eject
\pageno=1
\footline={\hss \folio \hss}

\line{\bf 1. Introduction\hfil}
\medskip\noindent
 Using an Effective Chiral Lagrangian [1] has been a popular approach
 to study the Symmetry Breaking Sector of the
 Standard Model in recent times. The reasons for this popularity
 are twofold. On the one hand, some calculations
 concerning the scattering of longitudinally polarized $W$'s and $Z$'s
 are greatly simplified in the limit where the Higgs mass is large
 if, instead of the full Standard Model, one
 just replaces these longitudinal degrees of freedom by the
 corresponding Goldstone bosons[2] and works with the self-interaction
 part of the scalar sector of the Standard Model. In the limit where
 the Higgs is very massive this self-interaction sector becomes
 a non-linear sigma model. Secondly, it was realized some time
 ago[3,4] that if one assumes a certain gap between
 the electroweak scale $G_F^{-1}$ and new resonances either scalar,
 such as the Higgs (in the minimal Standard Model),
 or vector-like, such as e.g. technirhos (in composite models), it
 was possible to parametrize in full generality the Symmetry Breaking
 Sector by the ${\cal O}(p^4)$ coefficients of an Effective Chiral
 Lagrangian describing the interactions of the Goldstone bosons
 associated to the breaking of the group $SU(2)_L\times SU(2)_R$ down
 to its diagonal subgroup.

We have examined the way these ${\cal O}(p^4)$ coefficients
are determined in the Standard Model and we have found a number of
difficulties in the way the problem is usually addressed. These
can be summarized as follows:  a) at what level one should
impose matching between the effective and fundamental theory
(S-matrix elements, light particle 1PI Green functions,
connected Green functions)? b) should one also consider
gauge non-invariant effective operators in the effective theory
since gauge invariance is lost in Green functions once the gauge is
fixed?

The experimental
 precision for some experiments sensitive to these coefficients
 is such[5] that the days of order-of-magnitude estimates are gone.
 It
 is  essential to be able to
 determine as precisely as possible the values of these ${\cal O}(p^4)$
 coefficients in the different models one chooses to compare with the
 experimental data. The differences are often minute, but this is where
 the clues as to what lies beyond the Standard Model are.

We have used the most recent available electroweak data, combined
with the preliminary determination of the top quark mass
by CDF, to set bounds on the values of the Effective Chiral
Lagrangian. The `oblique' corrections ---by far the dominant
ones--- turn out to be sensitive in a non-ambiguous manner
to {\it only one} combination
of coefficients of this effective theory. Our analysis, including
the leading two loop effects, is presented in the last section.
\bigskip

\line{\bf 2. Changing Variables in the Standard Model\hfil}
\medskip\noindent
In the usual representation of the Symmetry Breaking Sector
of the Standard
Model, the Goldstone bosons transform linearly under the $SU(2)_L\times
U(1)_Y$ gauge
group. However, if one is interested in a comparison with an
effective lagrangian approach where the Goldstone bosons
transform non-linearly it seems natural to implement
a change of variables in the Standard Model itself to
make the goldstones transform non-linearly too. In
this way the identification of fields and Green functions
will be clearer.

We write the Weinberg-Salam model
(without fermions) in the following form
$$
{\cal L}_{SM}=-{1 \over 2}{\rm Tr} W_{\mu\nu} W^{\mu\nu}
-{1 \over 4} B_{\mu\nu} B^{\mu\nu}+{1 \over 4}{\rm Tr} D_{\mu} M^{\dag}
D^{\mu} M -{1 \over 4} \lambda ({1 \over 2}
{\rm Tr} M^{\dag} M+{\mu^2 \over \lambda})^{2} \eqno(2.1)
$$
$M$ collects the scalar fields. Writing
$M=\sigma+i\vec{\tau}\vec{\omega}$ one recovers the
familiar lagrangian. Under the local $SU(2)_{L}\times U(1)_Y$ group
$M$ transforms as
$$M'(x)=e^{i {\vec{\alpha} \vec{\tau} \over 2}} M(x) e^{-i {\beta
\tau^{3} \over 2}}\eqno(2.2)$$
The covariant derivative $D_{\mu}$ acts on $M$ as
$$D_{\mu} M=\partial_{\mu} M + i g W_{\mu} M(x) -
ig' B_{\mu} M(x){ \tau^{3} \over 2}\eqno(2.3) $$
$B_{\mu}$ and $W_{\mu}={1 \over 2} W_{\mu}^i
\tau^{i} $ are the vector boson fields, and
$B_{\mu\nu}$ and $W^{\mu\nu}$ are the corresponding field strength
tensors.
The fields $W_{\mu}^{3}$ and $B_{\mu}$ are a combination
of the physical degrees of freedom $A_{\mu},Z_{\mu}$
$$\eqalign{
A_{\mu}     = W_{\mu}^{3} s_{w} + B_{\mu} c_{w} \qquad  \qquad
Z_{\mu}     = W_{\mu}^{3} c_{w} - B_{\mu} s_{w} }\eqno(2.4) $$
$c_{w}$ and $s_{w}$ being the cosinus and sinus of the Weinberg angle,
respectively. In the on-shell scheme, which we shall use,
$c_{w}\equiv M_{W}/M_{Z}$.

One should add to (2.1) the gauge-fixing and Faddeev-Popov terms
$$
{\cal L}_{GF}=-{1 \over 2\xi_{W}}\sum_{i=1,3} (\partial^{\mu}
W_{\mu}^{i}+ {i \over 4} g v \xi_{W}{\rm Tr}\tau^{i} M)^2
-{1 \over 2\xi_{B}}(\partial^{\mu} B_{\mu}-
{i \over {4 }} g^\prime v \xi_{B}{\rm Tr}\tau^{3} M)^2 \eqno(2.5)
$$
$$\eqalign{
{\cal L}_{FP}=& \partial^\mu c^{\dag}_{0} \partial_{\mu}
c_{0} +
 \partial^\mu c^{\dag}_{i} \partial_{\mu} c_{i}
 -{1 \over 8}
g^{\prime 2}
v \xi_{B} c^{\dag}_{0} {\rm Tr} M c_{0}
+g\, c^{\dag}_{i} (\partial^{\mu} W_{\mu}^{k} \epsilon ^{ikj} - {1
\over 8} g v \xi_{W} {\rm Tr} \tau^{i} \tau^{j} M) c_{j} \cr
&+{1 \over 8} \sqrt{g g^\prime}
\,v \, (g^\prime \xi_{B}
c^{\dag}_{0} c_{i}+g \xi_{W} c^{\dag}_{i} c_{0}) {\rm Tr} \tau^{3}
\tau^{i} M }
\eqno(2.6)$$
In
Landau gauge ($\xi=0$) ghosts decouple from
Goldstone bosons.

We are free to choose a different parametrization for the matrix $M$,
implying a field redefinition.
Not any transformation leaves the S-matrix elements between physical
states unchanged, however.
In the functional integral ${\cal Z}=\int {\cal D}
\phi e^{i {\cal S}+J \phi}$, where $\phi$
and $J$ collectively stand for
field and sources, respectively, such a field redefinition will
induce a jacobian. It was demonstrated in [6]
that if this jacobian
is the identity when  all fields are set to zero
then the corresponding transformation is an
allowed one. One of such allowed field redefinitions in the
Standard Model is precisely the
one mapping the linear onto the non-linear realization
$$
M=(v+\rho)U\qquad\qquad
U=\exp{i {{\vec{\pi}\vec{\tau}}\over v}} \eqno(2.7)
$$
where we already allow for a non-zero expectation value of the
matrix $M$ by introducing $v$. In the Standard Model,
$v=\sqrt{-\mu^2/\lambda}\simeq 250$ GeV.

We have
$$
\sigma=(\rho+v) \cos{{\pi \over v}}-v=
\rho -{1 \over 2v}\pi^{2}-{\rho \over {2v^2}} \pi^{2}+...\eqno(2.8)
$$
$$
\omega^{i}=(\rho+v)  {\pi^{i} \over \pi} \sin{{\pi \over v}}
=\pi^{i}+{\rho \over v} \pi^{i}-{1 \over {6v^2}}{\pi^{2}}
\pi^{i}+... \eqno(2.9)
$$
with $\pi=\sqrt{\sum_{i=1,3}}(\pi^i)^2$.
The jacobian of the change is
$$J(\rho,\pi^i)=\left\vert {{\partial (\sigma,\omega^i)} \over {\partial
(\rho,\pi^j)}}
\right\vert={1 \over v} (\rho+v)^3 {\sin^2{\pi \over v} \over \pi^2}
\eqno(2.10) $$
so indeed
$J(0,0)=1$, checking that the transformation is an allowed
one.

In fact, in the non-linear realization one expects
the covariant group measure
$d\mu(U)= \prod{d\pi^i}\sqrt{\det{g} }$ to appear. Here
$g$ is the group metric, given by
$$
g_{jk}={\delta^2 \over \delta(\partial_\nu \pi^{j}) \delta
(\partial^\nu\pi^{k})} ({v^2 \over 4}{\rm  Tr}
\partial_{\mu} U^{\dag} \partial^{\mu}U)
=v^2 {\sin^2{\pi \over v} \over \pi^2} (\delta_{jk}-{\pi_{j} \pi_{k}
\over \pi^2}) + {\pi_{j} \pi_{k} \over \pi^2} \eqno(2.11)$$
Working out $\sqrt{\det g}$ one finds
$$\sqrt{\det {g} }=v^2 {\sin^2{\pi \over v} \over \pi^2}\eqno(2.12)
$$
which is indeed contained in (2.10).
The jacobian (2.10) can be exponentiated
$$
\det{J}=\exp{\delta^{(4)}(0){\rm Tr}\ln{J}}\eqno(2.13) $$
In dimensional
regularization $\delta^{4}(0)$ is zero. But in other possible
regularization methods[7] this term will generate some tadpoles that
will be needed to yield a consistent result[8].
We will only use dimensional regularization here
and accordingly we will ignore
the jacobian altogether.

After substituting the parametrization (2.7) in (2.1)
one finds
$$\eqalign{
{\cal L}_{SM}=&{1 \over 2} \partial_{\mu} \rho \partial^{\mu} \rho
-\rho {\lambda v} (v^2+{\mu^2\over \lambda})
-{1 \over 2} \rho^2 (\mu^2 + 3 v^2 \lambda)
- \lambda v \rho^3 - {1\over 4}\lambda\rho^4 \cr
&+{1 \over 4} (\rho+v)^2 {\rm Tr}D_{\mu} U^{\dag} D^{\mu} U +
{\cal L}_{GF}^\prime
+ {\cal L}_{FP}^\prime}\eqno(2.14)$$
The primes in ${\cal L}_{FP}$ and ${\cal L}_{GF}$ indicate that
they also change under the  redefinition (2.7).

The couplings involving Goldstone bosons, collected in the
matrix $U$, have
changed completely when
compared to the linear realization.
All these couplings now involve derivatives and there are
an infinite number of them since we have a non-linear theory. Yet,
this non-linear theory is strictly equivalent to the Standard Model.
The couplings involving at least one gauge field remain unchanged
up to three fields, but for four fields and beyond
this is no longer the case. For instance, there is no vertex
$ W^{\mu +} W_{\mu}^{-} \pi^{+} \pi^{-} $.
Some new vertices
appear, e.g. $ \partial_{\mu} \pi^3 Z^{\mu}\rho^2 $,
and others change their coefficients.
The vertex that in the
old variables was ${i \over 2}g g^\prime s_{w} Z_{\mu} W_{\mu}^{+}
\omega^{-} \sigma$ now gets an extra factor of 2 and
becomes $ i g g' s_{w} Z_{\mu}
W_{\mu}^{+} \pi^{-} \rho$.
As befits a non-linear theory, we have vertices with
five, six, etc. fields, but they do not contribute to
the Green functions we will be interested in at the one loop level.
Of course, the different
coefficients have just the right values so as to render
a renormalizable and unitary theory as the Standard
Model should be.
\bigskip
\bigskip

\line{\bf 3. Heavy Higgs Limit\hfil}
\medskip\noindent
The parametrization (2.7) is particularly
useful in discussing the limit, within the minimal
Standard Model, when the Higgs particle is very heavy.
All the $M_{H}$ (or $\lambda$) dependence is
contained only in the propagator and
self-interactions  of the $\rho$ field,
while in the linear realization there are $\lambda$ dependences
in any scalar vertex.

The $\rho$ field itself interacts with the Goldstone bosons and
the gauge bosons through the operator
$O_{1}(x)= {\rm Tr}D_{\mu}
U^{\dag} D^{\mu} U$ and through the gauge-fixing and Fadeev-Popov terms.
The familiar 't Hooft
gauge-fixing term  (2.5) in the non-linear variables
reads
$$
{\cal L}_{GF}=-{1 \over 2\xi_{W}}\sum_{i=1,3} (\partial^{\mu}
W_{\mu}^{i}+ {i \over 4} g v \xi_{W}(v+\rho) {\rm Tr}\tau^{i} U)^2
-{1 \over 2\xi_{B}}(\partial^{\mu} B_{\mu}-
{i \over {4 }} g^\prime v \xi_{B}(v+\rho){\rm Tr}\tau^{3} U)^2
\eqno(3.1) $$
Unlike the usual formulation the Higgs field $\rho$ appears in the
gauge-fixing term.
Of course one could well have chosen some other gauge-fixing term,
since amplitudes are after all gauge independent, but, in practice,
we will be interested in making comparisons at other levels.
Most calculations in the Standard Model
are done in the 't Hooft gauge. In fact, the comparison
between theory and experiment is usually done for LEP physics
in the 't Hooft-Feynman gauge ($\xi=1$) [9,10]. `Observables' such
as the effective Weinberg angle $\bar{s}^2_w$ are actually
gauge dependent[13]. The $\rho$ couplings are simplest
in the 't Hooft-Landau gauge where
the Faddeev-Popov term does not lead to new Higgs
interactions and the additional
$\rho$ dependence appears only through the coupling
$$\rho (x)O_2(x)=-
 {iv \over 4}\rho (g\sum_{i=1,3}\partial^\mu W_\mu^i {\rm Tr}
\tau^i U -{g^\prime} \partial^\mu B_\mu {\rm Tr}
\tau^3 U)\eqno(3.2)$$
One can now formally perform
the functional integration over the Higgs field. The
result of such an integration will be a non-local
effective lagrangian of the type
$$
\int dk_{1} \int dk_{2} \dots \int dk_{n}
G_{\lambda}(k_{1},k_{2}, \dots ,k_{n}){\hat
O}(k_{1}){\hat O}(k_{2}) \dots {\hat O}(k_{n})
\eqno(3.3)$$
where $G_{\lambda}(k_{1},k_{2},\dots k_{n})$ are Green functions
that can be computed in a scalar field theory
involving only the $\rho$ field without any reference to the
gauge fields or Goldstone bosons.
These Green functions will depend on
$\lambda$ (or $M_{H}$) and an obviously important question is which is
the behaviour of $G_\lambda (k_1,k_2,\dots ,k_n)$ when
$\lambda\to\infty$. On general grounds we expect (3.3) to become
a local action in that limit.

A delicate point is whether one is allowed to take the $\lambda\to\infty$
limit directly in the
non-local lagrangian (3.3) and use the resulting local
effective field theory to compute quantum fluctuations for the remaining
fields[14]. Doing so would require the uniform
convergence of any momentum integral in which (3.3) is inserted,
a strong requirement that it is not always fulfilled.
After introducing the usual counterterms (e.g. using the on-shell
scheme[9-12] in the Standard Model) one is able to make all integrals
convergent, but typically they will be only conditionally convergent,
being the difference of two logarithmically divergent integrals.

In the on-shell scheme, if one considers
observables that depend only on renormalized self-energies (like the
celebrated
`oblique' corrections[9] contained in $\Delta r$, $\Delta\kappa$ and
$\Delta\rho$)
$$ \hat{\Sigma}(k^2)={\Sigma}(k^2)-{\Sigma}(M^2)+\delta Z_{2}
(k^2-M^2)
\eqno(3.4)$$
only one integral that depends on $M_H$ (actually $\sim$
$\log M_H$) and that is not uniformly convergent appears.
Thus
combinations that are finite in the large $M_H$ limit are combinations
where the potentially dangerous integral actually drops.
In these combinations (which, by the way, are the ones unambiguosly
predictable by an Effective Chiral Lagrangian[4,7]) we
are free to take the $M_H\to \infty$ limit {\it before} integrating
over the light degrees of freedom, simplifying the calculation
considerably.
This was the method used in [15] to determine the
contribution of the Standard Model to some LEP observables
in the large $M_H$ limit.

Unfortunately, it is not justified to retain only the leading
terms in the $1/M_H$ expansion of (3.3) for the three and four point
functions. In the on-shell scheme
the renormalized three and four point functions and their related
observables
will only be conditionally convergent, in general. We have to
keep the full non-local effective action (3.3).
(This point was overlooked
in [15], but subsequently realized in [16].)
To
determine the coefficients of the effective theory
reproducing the Standard Model we will
use the matching conditions, which will be
discussed in the next section.

It may be argued that one could expand the non-local action
in inverse powers of $M_H$ anyway provided that a physical
scale $\Lambda$ is introduced as cut-off. This is certainly
correct, since all integrals are then finite and well
behaved. Then one obtains a local action, equivalent to
(3.3) up to scales $k^2\sim \Lambda^2$ with some definite
values of the coefficients in this effective action. (These
coefficients, by the way, need not coincide with those
obtained by the use of the matching conditions, since the latter
contain some contribution from the light
particles.)
The above procedure, however, for a gauge theory is very difficult
to implement in an invariant way and we shall not pursue
this approach further here. Dimensional
regularization is the most useful regulator for gauge theories
and in dimensional regularization there is no manifest
decoupling of the heavy modes.
\bigskip

\line{\bf 4. Matching Conditions\hfil}
\medskip\noindent
We want to construct an effective
theory that reproduces the results of the Standard Model
without the Higgs. An obvious requirement this effective theory
should meet is to reproduce the same
S-matrix elements.
This is certainly a necessary condition, but it is not
the most useful way to proceed. For instance,
if we want to compare the fundamental and effective
theories at some intermediate steps (e.g.
at the level of the `oblique' corrections) we shall
need to deal with Green functions defined
off-shell both in the fundamental and in the effective
theory. Furthermore, to express these `observables'
in terms of the same set of parameters
($\alpha$, $M_W$, $M_Z$) it will be mandatory
 to renormalize both theories with the same conventions,
 requiring again off-shell Green functions.

It is often stated[17] that the matching can be done at the level
of the quantum effective action for the light fields;
that is, at the
level of the (light-fields) irreducible Green functions. This
is not quite correct. Let us see why.

{}From the generating functional $W=\log Z$
the renormalized connected Green functions of
the theory are obtained
$$
G_{c}(x_{1},x_{2},...,x_{n};\mu)={\delta^n W \over {\delta
J(x_{1}) \delta J(x_{2})... \delta
J(x_{n})}}\big\vert_{J=0}\eqno(4.1) $$
$\mu$ is some renormalization scale, chosen to be below
the mass of the particle we wish to integrate out, in
our case $\mu<M_H$.
Since the gauge fields have not been modified by the change
of variables (2.7),
the connected Green functions with only external gauge fields
should coincide when evaluated in the variables
($\sigma,\omega$) or ($\rho,\pi$).

This is not the case if one considers only one-particle
irreducible Green functions. The Legendre transform involves
the scalar fields too and these have changed.
For instance,
the 1PI Green functions
$\Gamma_{Z W W}$ and $\Gamma_{A W W}$
computed in the $(\sigma,\omega)$ or $(\rho,\pi)$ variables
at one loop change. In the large $M_H$ limit, the differences between
the linear $(\sigma,\omega)$ and non-linear $(\rho,\pi)$ realization,
denoted by $L$ and $NL$, respectively, are at one loop
$$\Gamma^{NL}_{ZWW}-\Gamma^{L}_{ZWW}=
 c_{w}{ 1\over 8} {g g'^2 \over {16 \pi^2}}(p_{2 \mu}g_{\lambda\nu}-p_{3
\nu}g_{\lambda\mu})\eqno(4.2)$$
$$\Gamma^{NL}_{AWW}-\Gamma^{L}_{AWW}=
- s_{w} {1\over 8} {g^3 \over {16 \pi^2}}(p_{2 \mu}g_{\lambda\nu}-p_{3
\nu}g_{\lambda\mu}) \eqno(4.3)$$
The culprit is one of the vertices containing four fields
that have changed in the non-linear realization.
(The full $M_H$ dependence of
these 1PI Green functions at one loop
in the linear realization can be found
in [10].) This simple
example should suffice to convince us that matching the Standard
Model to an effective theory by demanding the equality
of the 1PI Green functions is not correct, since a mere change
of variables (that does not affect the S-matrix elements)
in the Standard Model itself already changes these Green functions.

Of course
when we put everything together we must recover the same
connected functions with external gauge fields.
One can check this point easily for the connected
Green function  $\langle Z W^{+} W^{-}\rangle$ and $\langle A
W^{+} W^{-}\rangle$. To find the connected Green functions
one should add to Fig.1a the
reducible diagrams of Fig.1b.
It turns out that the 1PI Green function $\Sigma_{W\pi}$
also changes when we go to the $(\rho,\pi)$ variables; there
appears a new piece proportional to the squared momentum of the
internal $\pi$ field that cancels the $1/p^2_2$ of the propagator, yielding
a local contribution making up for the differences (4.2) and (4.3).

The matching conditions between an Effective Chiral Lagrangian
(ECL) and the Symmetry Breaking sector of the Standard Model (SM)
have therefore to be
imposed at the level of renormalized
connected Green functions for external
gauge fields (or on arbitrary S-matrix elements
between physical states, of course). We
will therefore tentatively demand that
$$
G_{\mu_1,\mu_2,\dots,\mu_n}^{SM}(x_1,x_2,\dots,x_n;\mu)=
G_{\mu_1,\mu_2,\dots,\mu_n}^{ECL}(x_1,x_2,\dots,x_n;\mu)
\eqno(4.4)$$
\bigskip

\line{\bf 5. Gauge Invariance and Matching Conditions\hfil}
\medskip\noindent
The effective low energy theory that one gets after
integrating out the heavy degrees of freedom in
the Standard Model, or, for that matter, in any theory with
the same local symmetry and the same
$SU(2)_L\times SU(2)_R\to SU(2)_V$ breaking
pattern, is the
gauged non-linear sigma model
$$
{\cal L}^{eff}=-{1 \over 2} {\rm Tr} W_{\mu\nu} W^{\mu\nu}
-{1 \over 4} B_{\mu\nu} B^{\mu\nu}+{v^2 \over 4} {\rm Tr} D_{\mu} U^{\dag}
D^{\mu} U + \sum_{i=0,13} a_i {\cal L}_{i}
+{\cal L}_{GF}+{\cal L}_{FP}
\eqno (5.1)
$$
A complete set of the operators ${\cal L}_i$ up to four derivatives
was given in [1].
Some of the operators are custodially
invariant, like those corresponding to the coefficients
$a_1$-$a_5$ (in the notation of [18,19] which we follow), while
others,
 such as $a_0$ and $a_6$-$a_{13}$, are not.
All of them are
gauge invariant operators.

This is a non-renormalizable theory but it may be rendered finite
at ${\cal O}(p^4)$
by redefining the $a_{i}
$ coefficients (only a few of them pick up divergent counterterms
actually[1]). The value
(at scale $\mu$) of these coefficients
is fixed by demanding the agreement with the Standard Model, thus
trading the dependence in the scale $\mu$ by $M_H$. In
the Standard Model therefore the bare coefficients are
of the form
$$ a_i={1\over {16\pi^2}}( c^{1}_i(C_{\epsilon}-
\log {M_H^2 \over \mu^2})+c^{2}_i)\eqno(5.2)
$$
with $C_\epsilon= {2 \over \epsilon} -\gamma + \ln 4 \pi$.
In another theory, $M_H^2$ is replaced by some other scale
$\Lambda$.

It is obvious that, in the Standard Model
after gauge-fixing the action, gauge non-invariant terms
are introduced.  In fact, the l.h.s. of the matching
conditions (4.4) is gauge dependent. Gauge non-invariant pieces
are also generated on the r.h.s. since in the effective
theory one needs to impose some gauge-fixing condition
as well. One might use gauge
conditions such as e.g.
$$ {\cal L}_{GF}^1=
-{1\over 2\xi_W}\sum_{i=1,3}
(\partial^\mu W_\mu^i +{i\over 4}gv^2\xi_W {\rm Tr}
\tau^i U)^2+\dots\eqno(5.3)$$
$$ {\cal L}_{GF}^2=
-{1\over 2\xi_W}\sum_{i=1,3}
(\partial^\mu W_\mu^i -{1\over 2}gv \xi_W \pi^i)^2
+\dots\eqno(5.4)$$
In fact, at the one loop level, for the Green functions
we are interested both are equivalent. At this point, it is
not obvious at all that the gauge non-invariant pieces that
are generated at one loop in the Standard Model with the usual
't Hooft-type gauge should be the same that appear from
a one-loop calculation with the pieces of
${\cal O}(p^2)$ in  ${\cal L}^{eff}$ using one of the
gauge-fixings  (5.3) or (5.4). Rather, one will
have to `fine tune' the gauge-fixing in the effective
theory to accomplish that. In other words,
one should also include  at ${\cal O}(p^4)$ some gauge non-invariant
operators on the r.h.s of the matching conditions. If not,
the matching conditions will overdetermine the coefficients
in the Effective Chiral Lagrangian and lead to inconsistencies.

This is an unwelcome complication.
Either we keep all BRS-invariant operators on the r.h.s of the
matching conditions
or we eliminate from
the Green functions to be matched the gauge dependent structures. Clearly
the latter is the simplest one and the one we take.
We shall therefore project the connected Green functions
on both sides of (4.4) on their tranversal components, by
multiplying them with the factor
$$ \prod_{i=1}^n (g^{\nu_i\mu_i}-{{k_i^{\nu_i} k_i^{\mu_i}}
\over k_i^2})\eqno(5.5)$$
This automatically eliminates all gauge-non invariant terms.
Of course even transverse parts may depend on the value
of the gauge parameter $\xi$. Fortunately, it can
be easily seen[19] that at the one loop level in the limit of
a large Higgs mass[12] all dependence on $\xi$ drops in
diagrams containing at least one Higgs internal line.
In the $M_H\to \infty$ limit the transverse projection of the
set of diagrams in the Standard Model that contains the
Higgs field; i.e. the set of diagrams whose contribution will
be implemented by the coefficients $a_i$ forms
a gauge invariant subset. One is then free to determine
these coefficients using, for instance, $\xi=0$ where the
chiral properties of the effective theory are manifest[1,18].
(The gauge-fixing term (5.3) respects the global
$SU(2)_L\times SU(2)_R$ invariance in this gauge.)

Let us now write explicitly the matching conditions
for the two point functions.
The renormalization constants for the fields and coupling
constants are defined in the usual way following the
on-shell prescriptions on both sides of (4.4).
The relevant diagrams will be those not common to
both sides of the equation and surviving the large Higgs
mass limit. They are discussed in [19]. We do not consider
tadpole
diagrams since they are exactly cancelled by redefining the second term in
(2.14).

$$ \eqalignno{
\Delta \hat{\Sigma}_{WW}=-
&{g^2 v^2 \over 4} (\Delta  Z_{\pi} -2
{\Delta
g \over g} -2 {\Delta  v \over v} ) - {g^2 \over {16 \pi^2}}({1
\over
8} M_{H}^2
-g^2 v^2 {3 \over 16} (C_{\epsilon} - \log {M_{H}^2 \over \mu^2}
+ {5 \over 6}))\cr
&+ q^2 (\Delta  Z_{W}+
{g^2 \over {16 \pi^2}} {1 \over
12} (C_{\epsilon} - \log {M_{H}^2 \over \mu^2}
+ {5 \over 6})) = 0  &(5.6) \cr  \cr
\Delta \hat{\Sigma}_{\gamma\gamma}=&q^2 (s_{w}^2 \Delta  Z_{W} + c_{w}^2
\Delta  Z_{B} -s_{w}^2 g^2 (a_{8}-2 a_{1})) = 0  &(5.7) \cr \cr
\Delta \hat{\Sigma}_{ZZ}=-
&{g^2 v^2 \over {4 c_{w}^2}}(\Delta Z_{\pi} -2 c_{w}^2 {\Delta  g
\over g} -2 s_{w}^2 {\Delta  g' \over g'} - 2{\Delta  v
\over v}+2 a_{0})\cr
&-{g^2 \over {16 \pi^2 c_{w}^2}}({1 \over 8}
M_{H}^2
-{g^2 v^2 \over c_{w}^2} {3 \over 16} (C_{\epsilon} - \log {M_{H}^2 \over
\mu^2}
+ {5 \over 6}))\cr
&+q^2 (c_{w}^2 \Delta  Z_{W}
+s_{w}^2 \Delta  Z_{B}
-c_{w}^2 g^2 a_{8} -2
s_{w}^2 g^2 a_{1} -(g^2 + g'^2)a_{13} \cr
&+  {g^2 \over {16 \pi^2 c_{w}^2}} {1 \over
12} (C_{\epsilon} - \log {M_{H}^2 \over \mu^2}
+ {5 \over 6})) = 0  & (5.8)\cr} $$

$$
\Delta \hat{\Sigma}_{\gamma Z}=-
gg'{v^2 \over 4} ({\Delta  g' \over
g'}-{\Delta  g
\over g}) + q^2 s_{w} c_{w} (\Delta  Z_{W}-\Delta  Z_{B} -
g^2 a_{8})
+q^2(c_{w}^2-s_{w}^2) gg'a_{1}=0   \eqno (5.9)
$$
where $\Delta$ means the difference between any quantity (self-energy,
renormalization constant) evaluated in the Standard Model
minus the same
quantity evaluated in ${\cal L}^{eff}$.
Since we work in the on-shell
scheme
and the renormalization constants of both theories have been generated
by the same renormalization conditions\footnote{$^1$}{In the
effective theory it is unnatural to demand
${\hat \Sigma}^\prime_H(M_H^2)=0$
as is usually done in the on-shell scheme,
since the Higgs is integrated out. It would be better to demand
a similar condition on the $\pi$ fields. However this does not affect
the present calculation}
we know [9] that they can be expressed
in terms of the
unrenormalized self-energies. One can calculate
their difference in both theories
$$ \eqalign{
\Delta  Z_{W}=&{1 \over s_{w}^2} (s_{w}^4-c_{w}^4) g^2
(a_{8}+a_{13}) -2 g^2 (a_{1}+a_{13})
-2 {c_{w}^2 \over s_{w}^2} a_{0}+
{g^2 \over {16 \pi^2}} {5 \over 6} (C_{\epsilon}-\log{M_{H}^2 \over
\mu^2}+{5 \over 6}) \cr
\Delta  Z_{B}=&g^2 (a_{8}+a_{13}) +2 a_{0}
-{g'^2 \over {16 \pi^2}} {5 \over 6} (C_{\epsilon}-\log{M_{H}^2 \over
\mu^2}+{5 \over 6}) + g'^2 a_{13} \cr}\eqno(5.10)$$
$$
\Delta  g  =\Delta  g'=0
$$
Putting  together (5.6) to (5.10)
one can determine some  coefficients
$$\eqalign{
&a_{0}={g'^2 \over 16 \pi^2} {3 \over 8} (C_{\epsilon}-\log{M_{H}^2
\over \mu^2} + {5 \over 6}) \cr
&a_{1}+a_{13}={1 \over {16 \pi^2}}{1 \over 12}
(C_{\epsilon}-\log{M_{H}^2 \over \mu^2}+{5 \over 6}) \cr
&a_{8}+a_{13}=0 \cr}\eqno(5.11)$$
Repeating the same procedure for the three  and four point
Green functions one gets
$$\eqalign{
&a_{2}={1 \over {16 \pi^2}}{1 \over 24}
(C_{\epsilon}-\log{M_{H}^2 \over \mu^2}+{17 \over 6}) \cr
&a_{3}=-{1 \over {16 \pi^2}}{1 \over 24}
(C_{\epsilon}-\log{M_{H}^2 \over \mu^2}+{17 \over 6}) \cr
&a_{4}-a_{13}=-{1 \over 16 \pi^2}{1 \over 12}(C_{\epsilon}-\log{M_{H}^2
\over \mu^2} + {17 \over 6}) \cr
&a_{5}+a_{13}={v^2 \over 8 M_{H}^2} -{1 \over 16 \pi^2}{1 \over
24}(C_{\epsilon}-\log{M_{H}^2
\over \mu^2} - {27 \pi \over {2 \sqrt{3}}} + {79 \over 3}) \cr
&a_{6}-a_{13}=a_{7}+a_{13}=a_{9}=a_{10}=0 \cr}\eqno(5.12)$$ %
To determine $a_4$ to $a_7$ and $a_{10}$ we have used the Equivalence
Theorem [2,20-21]. See also [22].

These values basically, but not quite, agree with those obtained in
[19]. As we have discussed in this section we cannot determine
with two, three and four point
gauge Green functions
alone
the values of all coefficients $a_i$
in the effective chiral lagrangian. For
instance $a_1$, $a_8$ and $a_{13}$
always appears in the combinations $a_1+a_{13}$ and $a_8+a_{13}$
while $a_{11}$ and $a_{12}$ drop from all transverse structures
in the Green functions considered. Note that in the evaluation
of the functional integral at the order we are working it is
legitimate to use the equations of motion for the $\pi$ fields coming from
the ${\cal O}(p^2)$ operators in the ${\cal
O}(p^4)$ terms and, if so, ${\cal L}_{11}$ and  ${\cal L}_{12}$ vanish,
and
${\cal L}_{13}$ does not provide new independent structures, so things
really are as they should.
\bigskip

\line{\bf 6. Oblique Corrections\hfil}
\medskip\noindent
Although we are not able to determine with gauge Green
functions alone all
coefficients but only some combinations of them, these are precisely the
ones that enter the physical observables. For instance, we may
choose to parametrize possible departures from the Standard Model
predictions in terms of the quantities $\epsilon_1$, $\epsilon_2$
and $\epsilon_3$[23] (or $S$,$T$,$U$ [24]). Then, the contribution to
these quantities from the operators ${\cal L}_i^\prime$ s are
$$2a_{0}\to \epsilon_1 \qquad -g^2 (a_{8}+a_{13})\to\epsilon_2
\qquad -g^2 (a_{1}+a_{13})\to \epsilon_3\eqno(6.1)$$
One way to proceed is to parametrize the `universal' part of
the radiative corrections in terms of the $\epsilon_i$ (basically
combinations of self-energies) and use the experimental data
to set constraints on them. In the SM each of the $\epsilon_i$ takes a
well defined value and depends logarithmically on the top-quark mass.
Beyond the Standard model, the Effective Chiral Lagrangian is
non-renormalizable theory and some cut-off effects remain. The latter
can be traced using (5.2) and (6.1) and we can form
combinations that are cut-off independent, at least at the one loop
level.

We prefer however to set the discussion in terms of quantities which
are directly observable. LEP basically measures two quantities of
relevance in the present
discussion, namely the effective mixing angle $\bar{s}_w$, extracted
from the forward-backward asymmetry $A_{FB}$ through
$$ \bar{s}^2_w= {1\over 4}(1-g_{V}/g_{A})    \qquad     A_{FB}=
{3 \over 4}\left( {2 g_{V } g_{A } \over g_{V }^2+g_{A }^2}
\right)^2
\eqno(6.2) $$
and the
leptonic width
$$ \Gamma_l = {G_{F} M_{Z}^3 \over 6 \sqrt{2} \pi}(g_{V}^2+g_{A}^2)
\left ( 1+ {3 \over 4} {\alpha \over \pi} \right)
\eqno(6.3)$$ %
$\Gamma_l$ is proportional to $\rho_Z$. $\rho_Z$
parametrizes
the strength of the neutral current at $s=M_Z^2$ in the improved
Born approximation (at tree level $\rho_Z= 1$). In this sense is
the counterpart of the
more familiar $\rho$ parameter (defined at $s=0$)
$${\cal A}^{NC}(s=M_{Z}^2)= \rho_{Z} \sqrt{2} G_{F} M_{Z}^2 {{J_{\mu}
J^{\mu}
} \over {{s - M_{Z}^2+ i {s \over M_{Z}^2} M_{Z} \Gamma_{Z}}}}
\qquad
{\cal A}^{NC}(s=0)=- \rho \sqrt{2} G_{F}  J_{\mu} J^{\mu}
\eqno (6.4)$$
 The  experimental data [5], assuming lepton universality
and correcting for the $\tau$ mass, gives
$A_{FB} = 0.0170 \pm 0.0016   $, $\Gamma_l= 83.98 \pm 0.18$ MeV,
which translates
into $g_{A}^2=0.25123 \pm 0.00056$, $g_{V}^2=0.00144 \pm 0.00014$
and $\bar{s}^2_W= 0.23107 \pm 0.00092$.
Given $m_{top}$ and $M_H$, the
SM value is just a dot in the
($\bar{s}^2_w,\Gamma_l$) plane.
In any effective theory we
have instead a line of points obtained by varying the (unknown) value
of the cut-off.
While we are unsure what value to
take for it, the line itself is unambiguously predicted
by one-loop chiral
perturbation theory, independently of the regulator one
uses[7]. Finding out which self-energies contribute
to $g_A, g_V$ we see that, in agreement with [4],
at the one loop level, we are sensitive to only
one combination of coefficients in the Effective Chiral Lagrangian,
namely
$$ L= -{2 \over 9} c_{w}^2 a_{0} + g^2 s_{w}^2 (a_{1}+a_{13})
+g^2 c_{w}^2 (a_{8}+a_{13})
\eqno(6.5)$$ %
{}From (5.11-12), in the Standard Model $L=0$[4].

With a relatively heavy top, such as
the one preliminarily reported in [25], $m_{top}=174 \pm
10^{+13}_{-12}$ GeV) some two-loop corrections are known to be
important.
Sizeable
contributions originate from a few genuine two-loop diagrams
that yield a $m_{top}^4$ dependence[26] and from the iteration of one
loop corrections through resummed propagators. The former, although
not negligible, give a small contribution for our purposes.
We have examined the contribution from the Effective Chiral
Lagrangian to the resummed propagators checking that $a_{11}$,
$a_{12}$ and $a_{13}$ still drop from the observables. Notice
that with a two loop precision we are not entitled to appeal to
the equations of motion derived from the ${\cal O}(p^2)$ terms. The
set of points obtained by varying $\log M_H$ (or $\Lambda$ in the
effective theory) is no longer a straight line, but deviations
are really not perceptible.

The results are shown in fig. 2.
We have plotted in addition of
$L=0$ (the value in the Minimal Standard Model) lines for the
values
$L= -{e^2 / 12 \pi^2}$ and $L=-{e^2 / 6 \pi^2}$. These
correspond to theoretical estimates[27] for $L$ in one-generation
technicolor models
with $N_{TC}=2, 4$, respectively. The same estimates in QCD
would give values  which are  between 30\% and 40\% below the
experimental results,
so we regard those as {\it lower bounds}. (A fact that can be rigorously
established in the large $N_{TC}$ limit.)
These lines do not agree in slope
with the one presented in [28].

Unless one is willing to make somewhat uncertain extrapolations
from QCD, the model with $N_{TC}=4$ and one full generation
of technifermions is not quite excluded at the
99\% confidence level, a conclusion that somehow runs contrary
to a widespread belief. A model with $N_{TC}=2$ falls within
the 68\% c.l. boundary. Even allowing for somewhat larger
values for $L$ it is hard to convincingly exclude this
model at this point.

In conclusion, we have revised thoroughly the procedure by means
of which one determines the coefficients in an Effective Chiral
Lagrangian that reproduce the Standard Model in the large
$M_H$ limit. We have pointed out a number of subtleties regarding
gauge invariance, commutation of limits and the precise
formulation of the matching conditions. At the end, we can
determine all experimentally relevant coefficients (but not
{\it all} coefficients). We have included the leading two loop
corrections to take a fresh look at the issue whether LEP
data really excludes technicolor models or not with the current
level of precision and the preliminary determination
of $m_{top}$.

\beginsection{\bf 8. Acknowledgements}

 We thank M.J.Herrero and H.Leutwyler for discussions that have
 triggered different parts of this work. The thank very specially
M.Martinez who has helped us in the analysis of the experimental data.
We acknowledge the financial support from CICYT grant AEN93-0695 and
CEE grant CHRX CT93 0343. J.M. acknowledges a fellowship from Ministerio
 de Educacion y Ciencia. D.E. would like to thank C.Garcia-Canal and
the Theory Group at Universidad de La Plata for
 the hospitality extended to him.

\beginsection{\bf References}

 \item{[1] }{
T.Appelquist and C.Bernard, Phys. Rev. D22 (1980) 200;
A.Longhitano, Phys. Rev. D22 (1980) 1166;
A. Longhitano, Nucl. Phys. B188 (1981) 118}
\item{[2] }{ M. Chanowitz and M.K. Gaillard, Nucl. Phys. B261 (1985)
379;  G.J. Gounaris, R. Kogerler and H. Neufeld, Phys. Rev. D34 (1986)
3257;
M. Chanowitz, M. Golden and H.Georgi, Phys.Rev. D36 (1987) 1490;
O.Cheyette and M.K.Gaillard, Phys. Lett B197 (1987) 205;
Y.P.Yao and C.P.Yuan, Phys. Rev. D38 (1988) 2237;
H.Veltman, Phys. Rev. D41 (1990) 2294
}
\item{[3] }{R. Renken and M.Peskin, Nucl. Phys. B211 (1983)
93; T.Appelquist, T. Takeuchi, M. Einhorn and L.C.R.
Wijewardhana,
Phys. Lett. B 232 (1989) 211;
A.Dobado and M.J.Herrero, Phys.Lett. B228 (1989) 495; J.F.Donoghue and
C.Ramirez, Phys. Lett. B234 (1990) 361;
B.Holdom and J.Terning, Phys. Lett. B247 (1990) 88;
A.Dobado and M.J.Herrero and J.Terron, Z.Phys. C50 (1991) 205, 465;
S.Dawson and G.Valencia, Nucl.Phys. B352 (1991) 27;
 M.Golden and L.Randall, Nucl. Phys. B361 (1991) 3;
T. Appelquist and G. Triantaphyllou, Phys. Lett. B278 (1992) 345;
J.Bagger, S.Dawson and G.Valencia, Fermilab-Pub-92/75-T,1992;
T. Appelquist and G-H. Wu, Phys.Rev.D 48 (1993) 3235}

\item{[4] }{A.Dobado, D.Espriu and M.J. Herrero, Phys. Lett. B255
(1991) 405}

\item{[5] }{Internal Note LEPEWWG/94-01 ALEPH 94-74 PHYSIC
94-63,DELPHI 94-33 PHYS 364 L3 Note 1599, OPAL Technical Note TN235 ,
May (1994)}
\item{[6] }{R.Haag , Phys. Rev. 112 (1958) 669;S.Coleman, J.Wess and
B.Zumino, Phys.Rev.177 (1969) 2239;C.G.Callan, S Coleman,
J.Wess and B.Zumino, Phys.Rev. 177 (1969) 2247}
\item{[7] }{D. Espriu and J.Matias, Nucl. Phys. B 418 (1994) 494}
\item{[8] }{J.Honerkamp and K.Meetz, Phys. Rev. D3 (1971) 1996;
G.Ecker and J.Honerkamp, Nucl. Phys. B 35(1971) 481;  52(1973) 211;
62 (1973) 509; T. Appelquist and C.Bernard, Phys. Rev D 23 (1981) 425}
\item{[9] }{G. Burgers and W. Hollik, in Polarization at LEP, CERN
Yellow Report, ed. G. Alexander et al. (CERN, Geneva, 1988);
M.Consoli and W.Hollik , in Z Physics at LEP1,
CERN Yellow Report, ed. G.Altarelli et al. (CERN, Geneva, 1989)
G.Burgers and F.Jegerlehner, ibid}
\item{[10]}{M.Bohm,H.Spiesberger and W.Hollik, Fortschr. Phys. 34
(1986) 687}
\item{[11]}{W.J.Marciano and A.Sirlin, Phys.Rev. D22 (1980)
2695, J.Fleischer and  F.Jegerlehner Nucl.Phys. B 228 (1983) 1}
\item{[12]}{K.I.Aoki, Z.Hioki,R.Kawabe,M.Konuma and T.Muta, Suppl. of
the Progress of Theoretical Physics 73 (1982) 1}
\item{[13]}{G. Degrassi and A.Sirlin, Nucl.Phys. B352 (1991) 342; P.Gambino and
Phys.Rev. D49 (1994) 1160}
\item{[14]}{
B.Ovrut and H.Schnitzer,Phys. Rev. D21 (1980) 3369;
Phys. Rev. D22 (1980) 2518;
Phys. Rev. D24 (1981) 1695 and references therein}
\item{[15]}{D. Espriu and M.J. Herrero,
Nucl. Phys. B 373 (1992) 117}
\item{[16]}{M.J.Herrero and E.R. Morales, Phys.Lett. B296 (1992) 397}
\item{[17]}{H.Georgi, in Proc. of the Workshop on Effective
Field Theories of the S.M., Dobogoko, Hungary, August 1991 (ed. U-G.
Meissner, World Scientific);
H.Georgi, L.Kaplan and D.Morin , Phys.Rev. D49 (1994) 2457}
\item{[18]}{F.Feruglio in Lectures at the $2^{nd}$ NATO Seminar,
Parma, Univ. di Padova (1992) DFPD92-TH-/50}
\item{[19]}{M.J.Herrero and E.R.Morales, Nucl.Phys. B418 (1994) 431}
\item{[20]}{A. Dobado and J.R. Pelaez  Preprint SU-ITP-93-33 (to be
published in NPB) and Phys.Lett. B329 (1994) 469; D. Espriu
and J. Matias, in preparation}
\item{[21]}{M.J.G. Veltman and F.J.Yndurain, Nucl.Phys. B325 (1989) 1;
S. Dawson and S. Willenbrock, Phys. Rev. Lett. 62 (1989) 1232}
\item{[22]}{M.J.Herrero and E.R.Morales, Preprint Universidad Autonoma
de Madrid, (1994) FTUAM 94/11 and FTUAM 94/12}
\item{[23]}{G.Altarelli and R.Barbieri, Phys. Lett. B253 (1991) 161;
G.Altarelli and R.Barbieri and S.Jadach, Nucl.Phys. B369 (1992) 3}
\item{[24]}{ M.Peskin and T.Takeuchi, Phys. Rev. Lett. 65 (1990) 964}
\item{[25]}{F.Abe et al. The CDF Collaboration FERMILAB Pub-94/097-E
CDF}
\item{[26]}{R. Barbieri et al., Phys. Lett. B288 (1992) 95;
J. Fleischer, O.V. Tarasov and F.Jegerlehner, Phys. Lett. B319 (1993) 249}
\item{[27]}{D.Espriu, E. de Rafael and J.Taron, Nucl.Phys.
B345 (1990) 22}
\item{[28]}{D.Buskulic et al.,The ALEPH Collaboration, Z.Phys. C60 (1993) 71}

 \vfill
\eject

\vfill\eject

\beginsection{\bf Figure Captions}

\bigskip\bigskip
\item{\bf Fig. 1.-} {a) One-Particle-Irreducible diagrams
that enter the $W^{+} W^{-} Z$ (or $A$) vertex. b)
Reducible
diagrams that restore the equality of the connected gauge
Green functions, as discussed in the text.}
\medskip
\item{\bf Fig. 2.- } {Plot of $\Gamma_l$, the leptonic width, versus the
effective mixing angle $\bar{s}^2_w$ for
 $m_{top}=174 \pm 10$ Gev and
100 GeV $\ge M_H \le$ 1500 Gev in the Minimal Standard Model (solid
line). The leading
two-loop corrections have been included. The dashed line (A)
corresponds to the same quantity (again including the leading two-loop
corrections) calculated in an Effective Lagrangian for $m_{top}=174$.
The agreement is
exact when $M_H\to \infty$. We then modify the coefficients of the
Effective Chiral Lagrangian to include QCD-like models for the
symmetry breaking sector with $N_{TC}\times N_D=$ 8 (B) and
16 (C). The values chosen correspond to theoretical
calculations that are really lower bounds in vector-like
models.
The elipsis corresponds to the experimental data with 68\% and
99\% C.L. The one-loop SM results are also shown (dotted line).}

\medskip
\vfill\eject

\bye